# PERFORMANCES OF NEUTRON SCATTERING SPECTROMETERS ON A COMPACT NEUTRON SOURCE

X. Fabrèges[1], A. Menelle[1], F. Ott[1], C. Alba-Simionesco[1]
[1] Laboratoire Léon Brillouin, CEA, CNRS, Université Paris-Saclay, CEA Saclay 91191 Gif sur Yvette France

N. Chauvin[2], J. Schwindling[2]
[2] IRFU/SACM, Université Paris-Saclay, CEA Saclay 91191 Gif sur Yvette France

A. Letourneau[3], A. Marchix[3],
[3] IRFU/SPhN, CEA, CNRS, Université Paris-Saclay, CEA Saclay 91191 Gif sur Yvette France

***Abstract.***

*There is currently a big effort put into the operation and construction of world class neutron scattering facilities (SNS and SNS-TS2 in the US, J-PARC in Japan, ESS in Europe, CSS in China, PIK in Russia). On the other hand, there exists a network of smaller neutron scattering facilities which play a key role in creating a large neutron scattering community who is able to efficiently use the existing facilities. With the foreseen closure of the ageing nuclear research reactors, especially in Europe there is a risk of seeing a shrinking of the community who would then be able to use efficiently the world class facilities. There is thus a reflection being conducted in several countries for the replacement of smaller research reactors with low energy accelerator based sources. We consider here a reference design for a compact neutron source based on existing accelerator components. We estimate the performances of various types of neutron scattering instruments built around such a source. The results suggest that nowadays state of the art neutron scattering experiments could be successfully performed on such a compact source and that it is thus a viable replacement solution for neutron research reactors.*

## 1 INTRODUCTION

As soon as neutron reactors were operating they have been used for neutron scattering experiments. Once the potential of neutron scattering was unraveled for magnetic studies and spectroscopic measurements dedicated facilities were built. The still world leading facility, the Institut Laue Langevin, started operation as soon as 1971. It is still operating the most performing instruments nowadays. There are currently about 50 nuclear research reactors operating across the world who are performing neutron scattering experiments [1]. Among these facilities, 20 are running a user program, that is, they are offering the possibility to academic users to perform neutron scattering experiments. During the 1980', a new type of neutron facilities based on the spallation reaction were developed. Eventually this led to the creation of 3 new facilities (IPNS [2], Los Alamos Neutron Science Center [3] and ISIS [4]) which were also running a user program. During the 2000', a second generation of spallation sources was built (SNS in the USA [5] and JPARC in Japan [6]). ESS [7] which can be considered as a third generation spallation source is currently being built in Europe. These last sources are very powerful and are able to replace or overtake nuclear reactors in terms of performances for neutron scattering. Considering the political situation in Europe, it is very unlikely that new nuclear research reactors will be built in replacement of the old ones. However, currently aging facilities are providing a broad user base to the most performing ones. It is thus necessary to try to find a solution to ensure that the broad user base (6000 users [8]) can be maintained in Europe in order to make the best use of the most powerful sources. The price tag of a full fledge spallation source is quite high (~1B€) and is difficult to bare by a single country. Hence any possibility to build neutron scattering facilities which could replace existing nuclear reactors would be welcome provided the investment is in the 100-200M€ range which would make it affordable to a single country on par with a synchrotron or a power laser facility. During the last decade a number of groups have independently considered the possibility of operating a high current / low energy proton accelerator to produce thermal and cold neutrons [9-10-11-12-13-14]. A few facilities have actually been built and are operating scattering instruments [15]. A UCANS network gathering these groups has been created [16].

In this communication we consider the design of a compact source based on existing components (proton source, accelerator, target, moderator, neutron optics) and evaluate the performances of a wide range of neutron scattering instruments which could operate on such a source. We focus mainly on elastic scattering instruments since the flux requirements are less stringent and the modelling of elastic instruments is significantly simpler than inelastic instruments.

## 2 REFERENCE DESIGN OF THE SOURCE

### 2.1 TARGET AND MODERATOR DESIGN

As a starting point we will consider in the following a source design based on existing components. We consider a source consisting of a proton source (100 keV, 60mA), a RFQ accelerator able to operate at proton currents up to 60mA bringing the proton energy to 3.6 MeV and a DTL section boosting the proton energy to 20 MeV. We also consider a pulsed operation (coupled with Time-of-Flight spectrometers) since this operation mode allows optimizing the (neutron production / energy input) ratio. Note that there are also accelerator based sources operating in continuous mode (PSI in Switzerland [17] and SARAF in Israel [18]). For the pulsed operation we consider a duty cycle of 4% which corresponds to a beam energy on the target of 50kW and an average proton current of 2.4mA. At the moment we do not make any assumption about the repetition rate and pulse length. Accelerator are very flexible so that an operation with 400µs long pulses at 100Hz or 2ms pulses at 20Hz could be considered. This shall be part of spectrometers optimization.

The corresponding accelerator components do exist. IPHI@Saclay for the RFQ and a section of the ESS DTL accelerator could be used to bring the proton beam to 20MeV with protons currents up to 60mA. The target would consist of a beryllium foil of about 1mm thickness which would convert the proton beam into fast neutrons. The choice of the target material is extensively discussed in [11]. However there are currently no Be target operating routinely at such power levels (>50kW) but various groups around the world are working on the issue especially for Boron Neutron Capture Therapy (BNCT) [19-20]. They are aiming at targets able to handle up to 300kW so that 50 kW seems a very reasonable goal. Note that it would be possible to operate at higher energies (such as 40MeV) so that the neutron flux would be multiplied by a factor 3.5. The 20MeV reference design limits the cost, the shielding and the activations issues.

The concept of a *compact neutron source* essentially lies in the fact that when operating at such low proton energies, the fast neutron source has a tiny volume (<0.01 liter) to be compared to 30 liters at ISIS. Hence the whole {target – moderator – shielding} assembly should be as compact as possible to maximize the neutron brilliance. On a compact source the active part of the moderator is less than 1 liter in volume (water or polyethylene) around which a reflector of about 10-15 liters should be added (PE, water). The configuration is thus of a very strongly coupled moderator geometry [9]. The use of hydrogenated materials allows to keep a compact reflector which limits the neutron pulse time broadening. It might be possible to use graphite or Be as a reflecting material to increase the neutron flux but this would lead to a broadening of the neutron pulse. Careful optimization of the target-moderator-reflector (TMR) geometry should be performed depending on the requirement for the neutron pulse structure.

## 2.2 SOURCE BRILLANCE

Figure 1a shows a basic non optimized design which illustrates the typical size involves in the TMR assembly. The corresponding neutron flux distribution is presented on Figure1b for the reference design (MCNP and GEANT4 Monte-Carlo simulations). For the instruments simulations we assume that we have a thermal (or cold moderator) providing a roughly Maxwellian spectrum. In order to estimate the moderator brilliance, rather than relying on numerical simulations, we use as a starting point the neutron flux value measured experimentally by Allen et al [21] at 10 MeV – 30µA on the Birmingham Nuffield Cyclotron using a moderator with a geometry almost identical to that of Figure 1a. The measured neutron flux (using gold foil activation) was $5\times10^8$ $n.s^{-1}.cm^{-2}.sr^{-1}$. Assuming that the flux scales as the proton current Gain1 = 60mA/0.03mA = 2000. The neutron yield increases when the proton energy is increased from 10 to 20 MeV (Gain2 = x4 as obtained from experimental data [22] and compatible with the values calculated using INCL4.6/ABLA07). The extrapolated brilliance value is thus $4\times10^{12}$ $n.s^{-1}.cm^{-2}.sr^{-1}$ (at 100% duty cycle) and $1.6\times10^{11}$ $n.s^{-1}.cm^{-2}.sr^{-1}$ (at 4% duty cycle). Note that these values are coherent with our recent measurements on IPHI at 3MeV, 10µA after appropriate scaling [23].

A source which the above brilliance was used as an input for the simulation of instruments using the McStas software [24] which is a tool dedicated to the modelling of neutron scattering instruments.

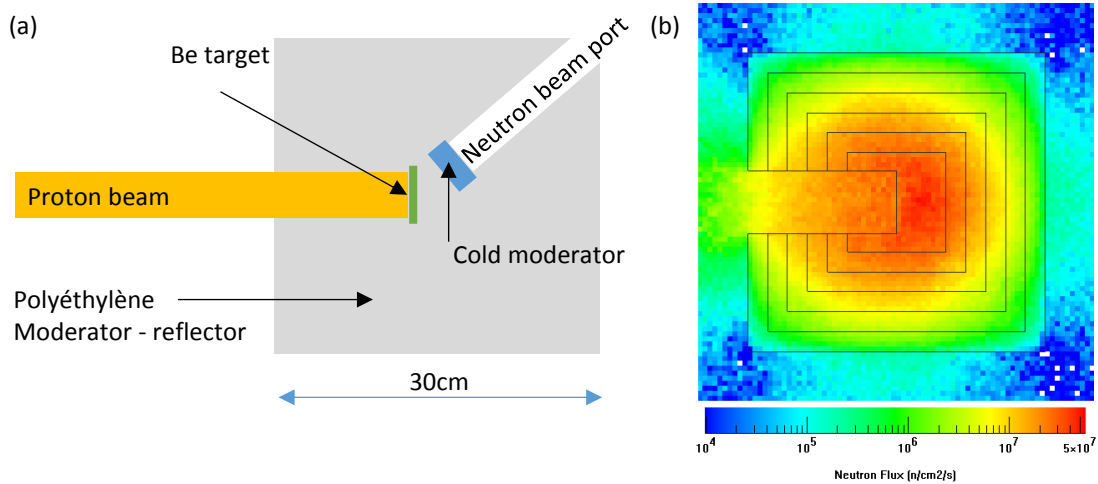

*Figure 1: (a) Sketch of a compact Target and moderator assembly. The target is a sheet of beryllium of thickness ~1mm and of surface 25-100cm². The moderator is a hydrogenated material (PE) which allows keeping short thermal neutron pulses. The assembly is surrounded by polyethylene, graphite or beryllium acting as reflectors. (b) Thermal neutron flux of such a source (without the cold moderator).*

## 3 INSTRUMENTS DESIGNS

In this section we will calculate the performances of instruments using the above TMR assembly. For each class of instruments we assume that the repetition rate and the pulse length can be freely optimized (within the 4% duty cycle power envelope). We consider that this makes sense since we are trying to evaluate the optimal performances of different types of instruments. Once such a source will be built, the details of the operation mode should be optimized with respect to the instruments most suited to the scientific case which the source should fulfill. Note also that it is likely that due to the limited costs, several TMR assemblies might be built around a single accelerator [11,25]. We also keep in mind that the aim is to build a not too expensive source so that the instruments should be kept as inexpensive as possible so as to be commensurate with the source price. We will also generally consider tow cases: (i) simple instruments easy to build, maintain and operate by non-expert users, (ii) instruments implementing new concepts but which usually require a lot more expertise to operate and data post-processing. These new concepts are not necessarily expensive but often require advanced tools for data post-processing to make it transparent to users. We hope that all the computing efforts put into ESS will make these new smart concepts accessible and transparent to casual users in the near future.

### 3.1 NEUTRON RADIOGRAPHY

Standard neutron radiography is a technique which is intrinsically handicapped on a pulsed source since it cannot benefit from the time structure of the source, since most of the current experiments are performed using a white beam. Nevertheless we consider such an instrument since it represents a significant share of nuclear reactor operation beam time (and provide a large part of the industrial income). The design of such an instrument being very simple it also allows introducing the basic concepts used for the calculations.

We consider a first instrument which would correspond to the G45 industrial radiography station at the Orphée reactor in Saclay (France) which is essentially dedicated to pyrotechnic component screening. The G45 instrument is installed at the end of a cold neutron guide (50m long) and the L/D of this instrument is 80. On our source, we thus consider an instrument which would have a length of 4 meters and a pinhole of size 5x5cm² (preceded by a 1meter neutron guide extracting the neutron from the moderator). With such a geometry, the flux at the detector position is $2 \times 10^7$ n/s/cm². The illuminated area is rather homogeneous over an area of about 25x25cm² which is perfectly suitable for industrial radiography using films or image plates. This flux is rather high but is obtained for a rather poor L/D ratio compared to state of the art instruments which are often operating at L/D of 400. The G45 instrument is using a Gd converter coupled with films for industrial radiography. In order to achieve high quality pictures with such a setup a fluence of $2 \times 10^9$ n.cm$^{-2}$ is used. Hence on SONATE, exposure times on the order of 100 seconds would be required which is perfectly acceptable. Note that significantly shorter exposure times are sufficient with other detectors (typically 20s with an image plate and even less with a scintillator + CCD system) but these systems have not yet been validated for industrial quality controls.

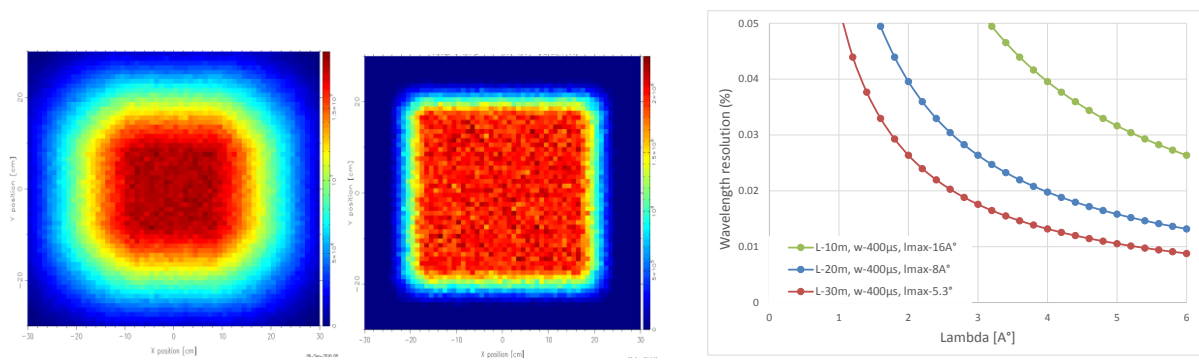

*Figure 2: (a) Illumination of the sample for an L/D = 80 radiography instrument. The flux in the central 20x20cm² part is on the order of $2 \times 10^7$ n/s/cm². (b) Illumination of the sample for an L/D = 240 radiography instrument. The flux in the central 40x40cm² part is very homogeneous but reduced to $2 \times 10^6$ n/s/cm². (c) Wavelength resolution for various instrument lengths at w = 400µs and f = 25Hz.*

For higher L/D ratios on the order of 250 as routinely used on the best radiography stations (ICON at PSI or CONRAD at HZB) the flux is divided by a factor 10 compared to the previous setup which gives a neutron flux of $2 \times 10^6$ n/cm²/s on our source. This should be compared to a flux on the order of $10^7$ n/cm²/s at ICON@PSI [26] or CONRAD@HZB [27]. Note that the latter instruments are routinely performing tomographic measurements so that a flux 5 times lower on our source would still be vastly sufficient for standard radiography (which is typically 100-200 times shorter to perform than tomography). Hence simple white beam radiography could be rather comfortably operated on our source. Such an instrument is also rather inexpensive to build.

There is currently an interest in developing time resolved radiography which provides a more detailed insight in the structure of metallurgical samples via a sensitivity of the neutron transmission to texture and strains as a function of the wavelength. The first such instrument to have been commissioned is RADEN@JPARC [28]. A second instrument IMAT@ISIS [29] is about to be commissioned. Other instruments (VENUS@SNS and ODIN@ESS) are in design phase. Energy resolved experiments can also be performed on continuous sources by using either velocity selectors or crystal monochromators. Such options are available at CONRAD@HZB and ICON@PSI. In the case of a continuous source, the energy resolution is obtained at the expense of a significant flux loss (depending on the energy resolution). For a wavelength resolution of 1%, the loss in flux is on the order of 200 [30]. On the other hand, on a pulsed source, provided high spatial resolution and time resolved detectors are available, the energy resolution is obtained for free by using the ToF technique. Note however that only the neutrons around the Bragg edge of interest are of interest so that the gain in efficiency does not simply scale by the previous factor between a continuous source and a pulsed source.

In order to optimize a ToF instruments, one has to set the aimed energy resolution and the useful wavelength range. In our case, the useful wavelength range is defined by the highest Bragg edge of the materials to be studied which is below 5Å for metals. The useful bandwidth thus ranges between 2 to 5 Å. We aim for an energy resolution of 1% which is smaller than typical Bragg edges widths (~2%). We thus have to choose a pulse width *w*, a repetition rate *f* and an instrument length *L* which match these requirements. Note that for a given energy resolution, the instrument length scales as the pulse width. At the detector position, the neutron wavelength is related to the neutron speed by the relation *λ* [Å] = *3956 / v* [m/s] = 3956 *t / L* where *t* is the time of flight. The wavelength resolution width is thus given by *Δλ = 3956 Δt / L = 3956 w / L*. The wavelength resolution is thus *Δλ/λ = (3956 * w) / (L * λ)*. Note that for very long instruments, the slower neutrons from pulse *n* might be overtaken by the fastest neutrons of pulse *n+1*. This should of course be avoided in the useful wavelength range. The largest usable wavelength (not subject to frame overlap) is given by $λ_{max}$ = 3956 / L * f. Figure 3 shows the wavelength resolution for *w = 400 μs* (which is the shortest pulse width which can be achieved using an efficient moderator), and increasing instrument lengths. For a given pulse width of *400 μs*, an instrument of about 30 m or more should be built to achieve a 1% wavelength resolution. Note that, in order to avoid frame overlap up to 5 Å, it is necessary to operate at a rather low frequency of 25 Hz which corresponds to a 1% duty cycle. For an L/D=500 and a wavelength resolution of 1%, the flux at the sample position on the order of $1 \times 10^5$ n/cm²/s which has to be compared to $1.5 \times 10^4$ n/cm²/s on NEUTRA@PSI and $5 \times 10^5$ n/cm²/s on ANTARES@FRM2 [29]. An energy resolved radiography station on our source might thus provide on-par performances with today best radiography facilities. Such an instrument would thus be rather expensive compared to the white beam radiography setup due to the long guide length. It might however be combined with a diffraction instrument (as IMAT@ISIS).

The key area in which instrumental progress will be made is neutron detectors. While industrial neutron radiography is still using photographic plates, all user facilities are currently using CCD cameras coupled with scintillators. While being vastly more flexible and efficient than films these systems suffer from the fact that as the spatial resolution is increased the thickness of the scintillator must be decreased and hence the detection efficiency drops. Current instrumental efforts have shown that it is possible to build very high spatial resolution detector (10-50 μm) with a very high detection efficiency together with a time resolution capability [31]. Such detectors could increase the performance of radiography stations by a factor 5 to 10 in terms of detection efficiency. Other methods using advanced optical elements [32] or coded source imaging [33] are also being implemented. These techniques while non trivial could boost the efficiency of neutron radiography stations by a factor 10 to 100.

## 3.2 REFLECTIVITY

Neutron reflectivity is a technique which is especially suited to Time-of-Flight measurements because reflectivity signals follow a dependence $R \sim 1 / Q^4 \sim \lambda^4$ while the neutron intensity in a Maxwellian distribution follows a $I(\lambda) \sim 1/ \lambda^4$ dependence. Hence the measured reflected neutrons R(Q). I($\lambda$) ~1 so that, to the first order, the reflectivity curve is measured at once with a constant statistics over the whole Q range. This has made neutron reflectivity a technique which can be very efficiently implemented on pulsed sources. ISIS is currently operating 5 reflectometers and even on reactors, all new reflectometers (FIGARO, PLATYPUS, HERMES) are using the ToF technique. It can be considered that wavelength resolutions $\delta\lambda/\lambda$ up to 10% are usable for reflectivity. For very thin layer of a few nanometers, a 20% resolution can even be used. HERMES@LLB is typically running with a 7% resolution (lowest possible resolution 95% of the time).

Let us assume a source operating with pulses of length 2 ms and with an operation frequency of 20 Hz. This corresponds to a duty cycle of 4% and is very close to the ESS neutron pulse structure. The wavelength resolution for various instrument length is presented on Figure 3. On a pulsed source the wavelength resolution degrades very quickly for short wavelengths. For pulse widths of w=2ms, the configuration (L=16m, w =2ms) is the most favorable: the wavelength band ranges from 3 to 12 Å which matches a cold neutron spectrum with a wavelength resolution of 12% at the peak flux (4 Å). The resolution improves quickly to 7% at 8 Å. A shorter instrument would degrade the wavelength resolution but would also increase the cut-off wavelength to 16 A° which might be usable in reflectivity experiments. A longer instrument (24 m) would provide a better wavelength resolution at the expense of a narrower wavelength bandwidth ($\lambda_{max}$ = 8 Å). While it seems acceptable to work in the (L=16 m, w=2 ms) configuration even though the resolution is as low as 15% at 3 Å, there are a number of situations where there is an interest in improving the resolution. Besides it is somewhat unsatisfactory to have a resolution varying from 15% to 4% over the bandwidth. A simple improvement such as a double disk-chopper [34] right at the moderator exit would (i) improve the wavelength resolution at short wavelengths (2) cut-off neutron pulse tails. Note that if the double disk chopper solution is used, due to the versatility of the system, a shorter instrument could be built (typically 12m) so that the resolution could be set at 10% below 6Å with the double disk chopper and improve down to 5% at 13Å. This would roughly match the POLREF@ISIS settings [35]. Figure 3 illustrates the operation principle of a double disk chopper and the wavelength resolution which can be achieved with such a system on a 12m long instrument.

In order to avoid a direct view from the moderator, since the instrument would be very short, a neutron mirror (e.g. m=6, length 500mm, height 20mm) could be used. The detector position (12m from the source) would be offset by more than 40cm from the direct view of the moderator.

We performed Monte Carlo simulations using McStas and the following design. From the moderator: straight guide of length 8m with m = 4, cross section 100x50mm²; a 2 m long collimator with F1 = 2 mm and F2 = 2 mm and a side guide with m = 4; a detector at 2 m from the sample position. The neutron flux at the sample position is $8 \times 10^6$ n/cm²/s which is on the order of CRISP@ISIS and HERMES@LLB. Hence it should be possible to operate a reflectometer with sufficient flux to perform useful science on our source.

Besides, it should be emphasized that new reflectometer concepts such ESTIA as being constructed at ESS [36] improve the efficiency of neutron reflectometers by a factor 30. Hence by implementing such a concept, it would be possible on a compact source to outperform currently existing instruments at medium reactor facilities.

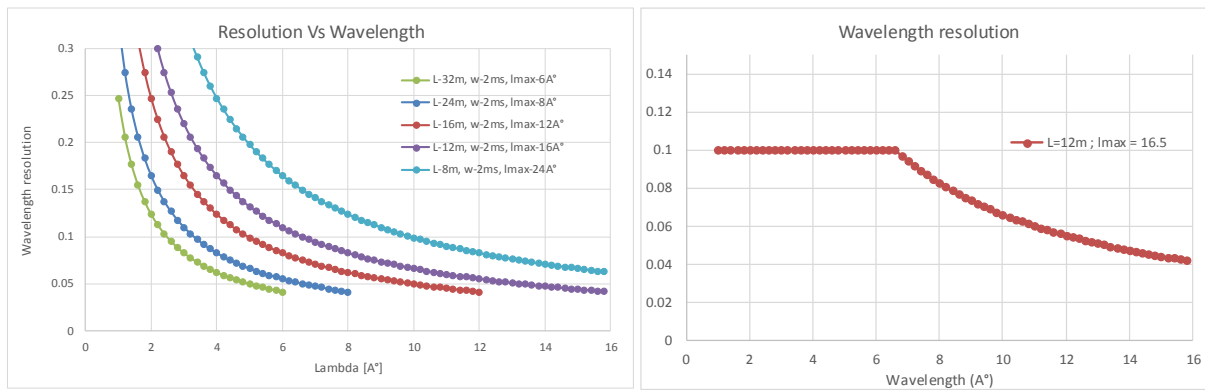

*Figure 3: (a) Wavelength resolution for various instrument lengths and a neutron pulse length of 2 ms. For a repetition rate of 20Hz, the longer the instrument, the shorter the frame overlap wavelength. (b) Wavelength resolution which can be achieved on a 12 m instrument with a double disk chopper.*

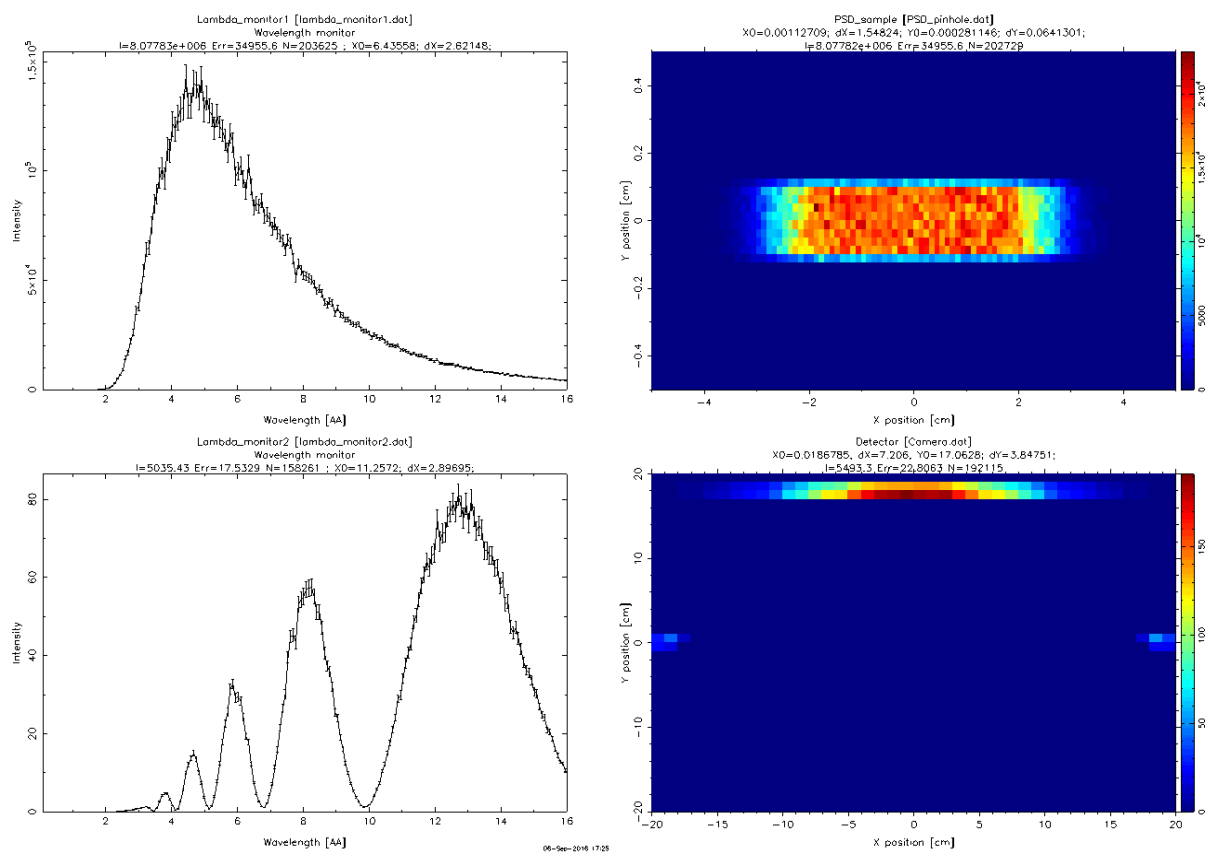

*Figure 4: McStas simulation of a 12m neutron reflectometer with a 10x5 cm² sample (Ni 20nm//Si) at an incidence angle of 3°. (a) Wavelength distribution incident on the sample. (b) Incident neutron distribution on the sample. (c) Wavelength spectrum measured in the specular reflectivity direction. (d) Neutron intensity on a PSD at 2 m from the sample. The top horizontal line corresponds to the reflected signal on the PSD detector*

## 3.3 SMALL ANGLE SCATTTERING

SANS scattering instruments are workhouse instruments for neutron scattering in soft matter and metallurgy. The instruments on the continuous sources are very productive because of the simplicity of the measurement procedure. Unfortunately, until now the implementation of Time-of-flight SANS instruments has not been very successful. While the construction of the instrument is rather straightforward and differs very little from a monochromatic instrument, complex data processing has hampered the very fruitful use of TOF-SANS instruments. Nevertheless, SNS has been successfully operating EQSANS for 5 years now, ISIS is building 2 instruments SANS2D and NIMROD and 2 SANS instruments will be built at ESS (LOKI and SKADI). Hence within a few years it is likely that most users will become familiar with TOF-SANS experiments and be able to properly exploit them. Considering the implementation of a SANS instrument on a CNS, we may use as a starting point the fact that SANS instruments are low resolution instruments which can be nicely operated with a wavelength resolution $\delta\lambda/\lambda$ ranging from 10 up to 20%. For SANS measurements, a wide lambda range is desirable to cover a Q range as large as possible. Hence a source with long pulses and a slow repetition rate is the most efficient way to operate. Let us again assume a source operating with pulses of length 2ms and with an operation frequency of 20 Hz (4% duty cycle). The wavelength resolution function is identical to the case of the reflectometer (Figure 3). Again, the configuration (L=12 m, w=2 ms) is rather favorable: (i) The wavelength band ranges from 3 to 16 Å which matches perfectly a cold neutron spectrum with a wavelength resolution of 16% at the peak flux (4 Å). The resolution improves quickly to 8% at 8 Å; (ii) for metallurgical studies, which require wavelengths above 5 Å, the resolution is below 12% which is optimal. A shorter instrument would degrade the wavelength resolution and the wavelength band extension beyond 16 Å is realistically not usable. A longer instrument would provide a better wavelength resolution at the expense of a narrower wavelength bandwidth. Note that in general the total length of a SANS instrument can be varied by moving the detector closer or further from the sample position. With PAXY@LLB, for example, the sample-detector distance varies from 1m to 7m and thus the total instrument length varies from 8 to 14m. Hence the instrument resolution varies as a function of the configuration. It might be useful to have a double disk chopper which allows tuning the pulse length even though this is not compulsory. We also recall that the above curves scales as *w/L*. Hence the same resolution functions are obtained for a pulse twice as long and an instrument twice as long.

We suggest that an instrument with the following specifications would be perfectly suitable for SANS studies: (i) cold source; (ii) Source-Sample distance of 8 m. Sample – detector distance variable from 1 to 7m, total flight path ranging from 9 to 15m; (iii) useful bandwidth from 3 Å to 16 Å (depending on the total instrument length). If one compares with the SANS2D@ISIS (f = 10 Hz, Source-sample = 19 m, Total length = 21 – 31m / $Q_{min}$~0.002 Å$^{-1}$, $Q_{max}$~3 Å$^{-1}$), the above specifications scale exactly as the operating frequency. At 20 Hz, an instrument should be twice as short as an instrument operating at 10 Hz for the same Q range.

Monte-Carlo simulations for various configurations were performed. The results are summarized in Table 1. The flux are also exactly on par with the PAXE instrument at the LLB [37]. The calculated flux are also on the order of the flux expected on SANS2D at ISIS TS2 (~$10^6$ cm$^{-2}$ s$^{-1}$) [38].

| Configuration | $L_g$ (m) | $L_1$ (m) | $L_2$ (m) | $L_{tot}$ (m) | $D_1$ (mm) | $D_2$ (mm) | Flux (n/cm²/s) |
|---|---|---|---|---|---|---|---|
| Small Q | 1 | 7 | 7 | 15 | 20 | 16 | $1\times10^6$ |
| Medium Q | 4 | 4 | 4 | 12 | 20 | 16 | $3\times10^6$ |
| High Q | 6 | 2 | 1 | 9 | 20 | 16 | $9\times10^6$ |
| PAXE [36] | 6 | 2.5 | 2.5 | 11 | 12 | 8 | $2.6\times10^6$ ($0.7\times10^6$ on PAXE) |

*Table 1: Flux at the sample position for various SANS configurations (small, medium, high Q). $L_g$ being the length of the guide from the source to the first pin-hole, $L_1$ being the length of the collimation and $L_2$ being the sample-detector distance. $D_1$ and $D_2$ are the sizes of the holes at the entrance and exit of the collimator.*

A few possibilities are available to increase the brilliance of SANS instruments. The simplest to implement is the multi-slit set-up which may increase the efficiency of the instrument by a factor 5 to 10 at the cost of a more complex data post-processing treatment. Due to the rather high brilliance of SANS instruments, multi-slit setups are not used on SANS instrument (even though they have been routinely used on laboratory SAXS instruments for decades)

## 3.4 POWDER DIFFRACTION AND SINGLE CRYSTAL DIFFRACTION

Power diffraction instruments are also workhorses in neutron diffraction. High resolution instruments are used for structure determination. High flux instruments are used to study phase transitions. Efficient powder diffraction in Time-Of-Flight mode requires a large coverage of the space with detectors which turns into a rather expensive instrument (contrary to monochromatic instruments). The same comment applies to single crystal diffraction, an efficient spectrometer can be built only with a large space coverage (~$2\pi$ steradians).

We will consider a design for a powder spectrometer aiming at replacing G41@LLB (low resolution - high flux). The G41 instrument at the LLB [39] is operating with a graphite monochromator on a cold neutron guide. The horizontal divergence is 0.3° and the vertical divergence is 3°. The detector is covering a solid angle of 80° x 3°. The neutron flux at $\lambda$ = 2.43 Å is $4\times10^6$ n/cm²/s at the sample position.

We consider the following design for a powder diffractometer (PRESTO): (i) cold source, (ii) pulse width w = 250µs, f = 40 Hz (1% duty cycle), (iii) Source-sample distance 52m, (iv) sample-detector distance of 1m, total flight path of 53m. This leads to a usable bandwidth ranging from 1.4Å to 3.3Å with $\Delta\lambda/\lambda$ ranging from 1.3% to 0.5%. The horizontal and vertical divergence are set to 0.6°. The flux on the sample would be $2\times10^6$ n/cm²/s but the operation in ToF mode provides an extra efficiency so that with the G41 detector the PRESTO configuration would offer performances on the order of 70% these of G41. However, a key advantage of ToF powder diffraction instruments is that they naturally offer a good vertical divergence due to the absence of focusing monochromator so that any gain in the detection solid angle coverage translates into a net gain in the instrument performances. For example, by using the 7C2@LLB detector which is covering a solid angle of 120°x20° with a detection efficiency of 90% a gain of a factor 20 would be obtained at no cost in performances. This shows that a powder instrument on our source can easily outperform existing very productive instruments at neutron reactor facilities.

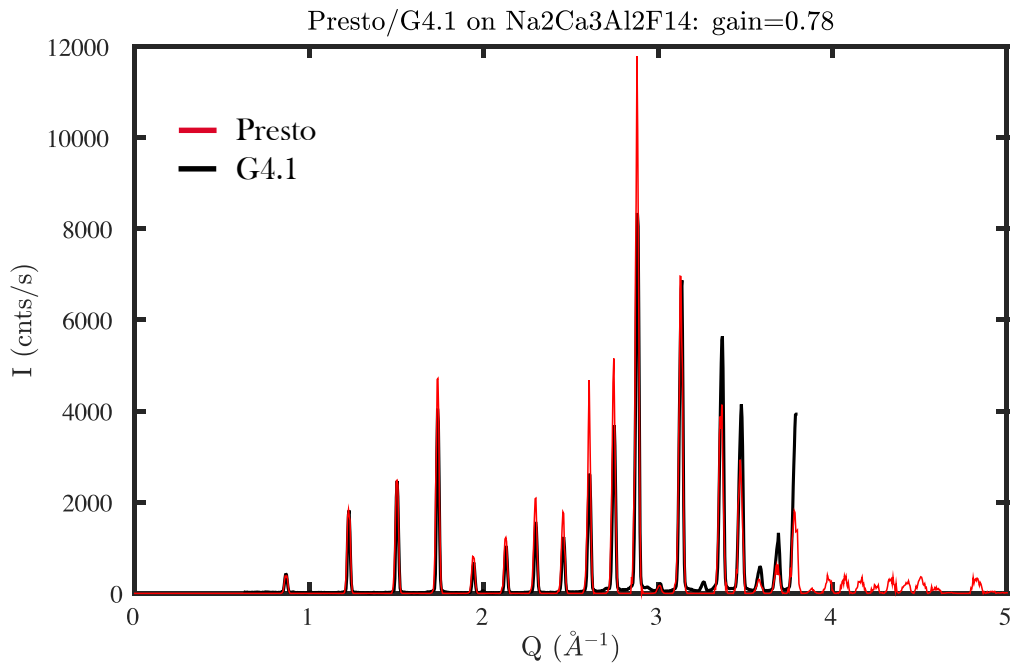

*Figure 5: Comparison of the diffraction spectra measured on G41 ($\lambda$ = 2.43Å) and PRESTO (ToF). A wider Q range is accessible in ToF (above 4Å$^{-1}$).*

## 3.5 SPECTROSCOPY

While two thirds of the neutron scattering experiments use elastic scattering, inelastic scattering is one of the great strength of neutron scattering thanks to the very low kinetic energy of neutrons (in the meV range). This allows for the study of excitations with energies ranging from neV (with spin-echo techniques) up to 0.1eV using Triple Axes techniques. However spectroscopic measurements are usually more flux demanding than elastic studies.

Nevertheless, it is possible to estimate the performances of some inelastic instruments on a compact neutron source. In the case of the Spin-Echo technique, very low wavelengths resolutions can be used (up to 20%) so that a very short instrument can be build. An instrument of 12 m on a 2 ms pulse length source with collimations suitable for spin-echo (~0.5°) would achieve a non polarized brilliance of about $1\times10^7$ n/cm²/s. Operating with polarized neutrons would lead to a loss of a factor 3 to 4, that is the flux at the sample position would be on the order of $3\times10^6$ n/cm²/s. This is significantly lower than the flux at MUSES@LLB [40] which is on the order of $2\times10^7$ n/cm²/s (at 5 Å). The spin-echo experiments are already demanding on existing reactor sources so that such a flux reduction would preclude any useful experiment on SONATE. This is heuristically confirmed by the fact that ISIS is not operating any regular Spin-Echo instrument. However, the Multi-MUSES project is aiming at boosting the efficiency of the MUSES spin-echo instrument by a factor 70. In this case, such an instrument on our source would provide a flux higher (x10) than existing state of the art spin-echo instruments but at the expense of a complex and costly instrument.

Another class of spectroscopic instruments are the so called Time-Of-Flight instruments which provide spectroscopic measurements in the 20μeV up to 1meV. Such instruments can be built in a direct (e.g. MERLIN@ISIS) or inverse geometry (e.g. OSIRIS@ISIS). In the case of the direct geometry configurations very large detectors areas (20-30 m²) are required which make these instruments very expensive. Inverse geometry spectrometers use crystals as energy analyzers. A rough estimate assuming an incident vertical and horizontal divergence of 1° on the sample and a 100 μeV energy resolution gives a flux of $6\times10^7$ n/cm²/s which is on par with the flux quoted for the OSIRIS instrument at ISIS. Such instruments seem perfectly viable on a compact neutron source.

The third class of spectroscopic instruments are the triple axis instruments which have been in operation since 1970 at reactor facilities and which allow a very controlled measurement at specific points in the reciprocal space. Currently there is no technical solution to match the performances of these instruments on a pulsed source. This is illustrated by the fact that no spallation source (ISIS, SNS, JPARC) is operating such instruments.

## 4   DISCUSSION

We have considered various classes of neutron scattering instruments built on a compact neutron and compared their performances to existing instruments at reactor or spallation facilities. The results are summarized in Table 2. From our simulations, assuming conservative designs (reasonable heat load on the target, non-optimized moderator, simple instrument designs), we think that it should be possible to build a compact neutron source with performances in the range of ORPHEE / ISIS TS2 for a cost on the order of M€100. Such a source (with a single target) could accommodate about 5 instruments. The possibility of setting up a second target using beam multiplexing would be welcome in order to accommodate more instruments.

| Technique | Flux on sample | Reference spectrometer | Potential gains |
|---|---|---|---|
| Imaging (white beam) | $2 \times 10^6$ n/s/cm² (for L/D = 240) <br> $2 \times 10^7$ n/s/cm² (for L/D = 80) | ICON@PSI $1 \times 10^7$ n/s/cm² <br> CONRAD@PSI $1 \times 10^7$ n/s/cm² (for L/D = 240) | MCP detectors x5 <br> Coded Source Imaging x10 |
| Imaging (time resolved) | 1E5 n/s/cm² (for L/D = 500) $d\lambda/\lambda = 1\%$ | NEUTRA@PSI $1.4 \times 10^7$ n/s/cm² (1%) <br> ANTARES@FRM2 $5 \times 10^5$ n/s/cm² | MCP detectors x5 <br> Coded Source Imaging x10 |
| SANS | $1 \times 10^6$ n/s/cm² (low Q) <br> $3 \times 10^6$ n/s/cm² (med Q) <br> $9 \times 10^6$ n/s/cm² (high Q) | PAXE@LLB (low Q) $0.7 \times 10^6$ n/s/cm² <br> SANS2D@ISIS $1 \times 10^6$ n/s/cm² (under construction) | Slit setup x10 <br> Focusing optics for VSANS (small Q) x10 |
| Reflectivity | $8 \times 10^6$ n/s/cm² | HERMES@LLB $8 \times 10^6$ n/s/cm² <br> POLREF@ISIS $\sim 1 \times 10^7$ n/s/cm² (under construction) | SELENE concept x10 <br> Advanced Deconvolution x3 |
| Spin-Echo | $3 \times 10^6$ n/s/cm² | MUSES@LLB $2 \times 10^7$ n/s/cm² (at 5A°) | Multi-MUSES (x70) |
| Low resolution powder diffraction | $2 \times 10^6$ n/s/cm² | G41@LLB $2 \times 10^6$ n/s/cm² | Large solid angle detector (7C2 type) x20 |
| TOF | $6 \times 10^7$ n/s/cm² | OSIRIS@ISIS $3 \times 10^7$ n/cm²/s | |

*Table 2: Comparison of the performances of various types of instruments on a compact source with existing instruments.*

While the performances of such a source would not match the performances of ESS, it would still serve the need of about 50-75% of the users who do not need extreme neutron flux. Such sources could successfully replace aging neutron reactors as well as lower the barrier for new entrant countries in the field of neutron scattering.

The current design is conservative and is using existing components (accelerator, moderator, target). Further brilliance gains could be obtained:

- The duty cycle has been set to 4% with a maximum current of 60 mA (ESS settings) With a modified design, it is possible to increase the duty cycle or the maximum current by a factor 2. Proton beams with currents of 100mA and more are foreseen for several on-going project.
- We used the neutron output from the most basic moderator geometry. A gain of a factor 2-3 is very likely if a 1D moderator geometry is considered. Very significant gains (x3) have been obtained for the neutron brilliance of the ESS moderator by optimizing its geometry.
- Very simple instruments have been considered (low cost – easy operation). More complex instruments could lead to significant gains (x2 to x10).
- The proton beam energy can be boosted to higher values (40-60MeV) leading to gains in neutron production of a factor (x 3.5 to x 6.5)

Hence, it is very likely that compared to the above table, instruments which are 10 times more efficient could be built in the future.